\documentclass[runningheads,a4paper]{llncs}

\usepackage{amssymb}
\usepackage{graphicx}

\usepackage{url}
\urldef{\mailsa}\path|mah.krantz@gmail.com,|
\urldef{\mailsb}\path|{johan.linaker, martin.host} @cs.lth.se,|
\newcommand{\keywords}[1]{\par\addvspace\baselineskip
\noindent\keywordname\enspace\ignorespaces#1}

\begin{document}

\mainmatter  

\title{Investigating applicability of inner source on small development teams}

\titlerunning{Investigating applicability of inner source on small development teams}

\author{Maria Krantz, Johan Lin{\aa}ker, Martin H\"ost \\}
\authorrunning{Maria Krantz, Johan Lin{\aa}ker, Martin H\"ost \\}

\institute{Software Engineering Research Group, Computer Science, Lund University\\
\mailsb}

%
%

\toctitle{Lecture Notes in Computer Science}
\tocauthor{Authors' Instructions}
\maketitle

\begin{abstract}
The phenomenon of adopting open source software development practices in a corporate environment is known by many names, one being inner source. 
The objective of this study is to investigate how an organization consisting of small development teams can benefit from adopting inner source and assess the level of applicability. 
The research has been conducted as a case study at a software development company. Data collection was carried out through interviews and a series of focus group meetings, and then analysed by mapping it to an available framework. 
The analysis shows that the organization possesses potential, and also identified a number of challenges and benefits of special importance to the case company. 
To address these challenges, the case study synthesized the organizational and infrastructural needs of the organization in a requirements specification, describing a technical infrastructure, and a suitable organizational context and work process.

\end{abstract}

\section{Introduction}

Many open source software products have been successful in recent years, which have led to an increased interest from the industry to investigate how the development practices could be introduced in a corporate environment and take advantage of the benefits seen in open source projects. Such practises include e.g. universal access to project artefacts \cite{Lindman08}, early and frequent releases, and ``community'' peer-review \cite{Gurbani06}.

Mistrik et al.\ \cite{Mistrik10} address how closed development organizations could benefit from open source practices as an area where further research is needed. Though studies conducted so far are quite limited, several success stories \cite{Wesselius08}, \cite{Gurbani06}, \cite{Dinkelacker02}, \cite{Lindman08}, \cite{Riehle09} can be found of large corporations adopting open source development. 

The phenomenon of adopting these development practices in a corporate environment has in research been called \emph{inner source} \cite{Stol11}, \cite{Gaughan09}, \emph{community source} \cite{Taft09}, \emph{corporate open source} \cite{Gurbani06}, \cite{Gurbani10} and \emph{progressive open source} \cite{Dinkelacker02}. 
In this report we have chosen use the term inner source, as described by Stol et al.\ \cite{Stol11}. 

The changes required when adopting inner source in a corporate environment led Gurbani et al. \cite{Gurbani10} to suggest two different methods to effectively manage inner source assets; an infrastructure-based model and a project-based model.

In the infrastructure-based model, the corporation provides the critical infrastructure that allows interested developers to host individual software projects on the infrastructure, much like SourceForge\footnote{\url{http://www.sourceforge.com/}} or Github\footnote{\url{http://github.com}} does with open source projects. Platforms like these, also known as forges  \cite{Riehle09},can be resembled as a component libraries where each project represents a component of different abstractions, e.g. modules, frameworks or executables. Developers can browse between the components and use or contribute to those they wish.  The reuse of software can be considered opportunistic or ad hoc and there is no limitation on the number of projects to be shared within the organization. Success stories include cases from SAP \cite{Riehle09}, IBM \cite{Sabbah05}, HP \cite{Dinkelacker02} \cite{Melian08} and Nokia \cite{Lindman08}.

In the project-based approach the software is managed in a project, instead of as a long-term infrastructure. Gurbani et al.\ \cite{Gurbani10} describe how an advanced technology group, or a research group funded by other business divisions in a corporation takes over a critical resource and makes it available across the organization. This team is often referred to as the ''core team'' and is responsible for the project and the decision making. Philips Healthcare \cite{Wesselius08} and 
Alcatel-Lucent \cite{Gurbani10} are two documented cases where this variant has been adapted.

In order to asses the applicability of inner source on an organization, Stol \cite{Stol11} developed a framework. This framework is based on reviewed literature and a case study of a software company referred to as ``newCorp''. Though the framework focuses on project-based models, it is based on success factors and guidelines described in both project-based and infrastructure-based case studies.
The framework consists of 17 elements divided into four categories; Software product, Development practices, Tools and infrastructure, and Organization and community. The elements can be found in the left column of Table \ref{SummaryChart}.

Adopting inner source requires significant effort and change management, which is one reason why it may be of interest to start on a smaller scale before investing globally. However, this requires an understanding of how inner source can be implemented on smaller teams and what parts that can be implemented and evaluated. 

This study is focused towards the latter and aims to contribute theoretically by using a new way of assessing the applicability of inner source, identifying key benefits and challenges as well as synthesising a solution that addresses the organizations needs.

The outline of this paper is as follows. In Section 2 the research methodology is presented and the results are presented in Section 3. The results are discussed and further analysed in Section 4, expected implications of an introduction of inner source is also discussed. The validity of the conducted research is presented in Section 5 and the research results are summarized in Section 6.

\section{Methodology}
This research is of a problem-solving nature and conducted as a case study with an exploratory strategy approach \cite{Runeson12} \cite{Yin02} \cite{Robson02}. The case company is experiencing problems in regards to its reuse of code and overall efficiency. The hypothesis is that the concept of inner source, as described earlier, can help the organization manage these issues. 

The organization was observed in order to further define the problem. Then it needed to be assessed whether inner source would fit the organizations or not, and what the challenges would be. Based on the findings, the parts of inner source suitable for the organizations needs were synthesized in a requirement specification describing a technical infrastructure together with an organizational context and work process.

This improving approach can be compared to that of action research. However, this study has focused on the initial parts and proposed a solution. This is yet to be implemented and evaluated. I.e. the complete change process is yet to be observed. Due to limitations in time for the researchers and organizational conditions of the case company, this is left for future research.

\subsection{The case company}
An international software development firm, hereby known as ``the case company'', has been chosen for this study. The case company has a division based in a local office in Sweden which specializes in rapid software development and deployment of projects where the customers seek a combination of high quality and a fast release.

The scope is limited to the local division, though a network of corresponding divisions is established globally. The division of interest is divided into two teams with similar set-up and structure. Each team consists of 20-25 engineers including developers, testers and project managers. 

Reuse today within the division is insufficient, which was one reason that the organization was interested in the topic. Knowledge of modules and functions developed within the projects are spread orally and physically in an unstructured manner which causes redundant work and loss of information.

\subsection{Case study steps}
\begin{figure}
\vspace{-5pt}
\begin{center}
\includegraphics[scale=0.65]{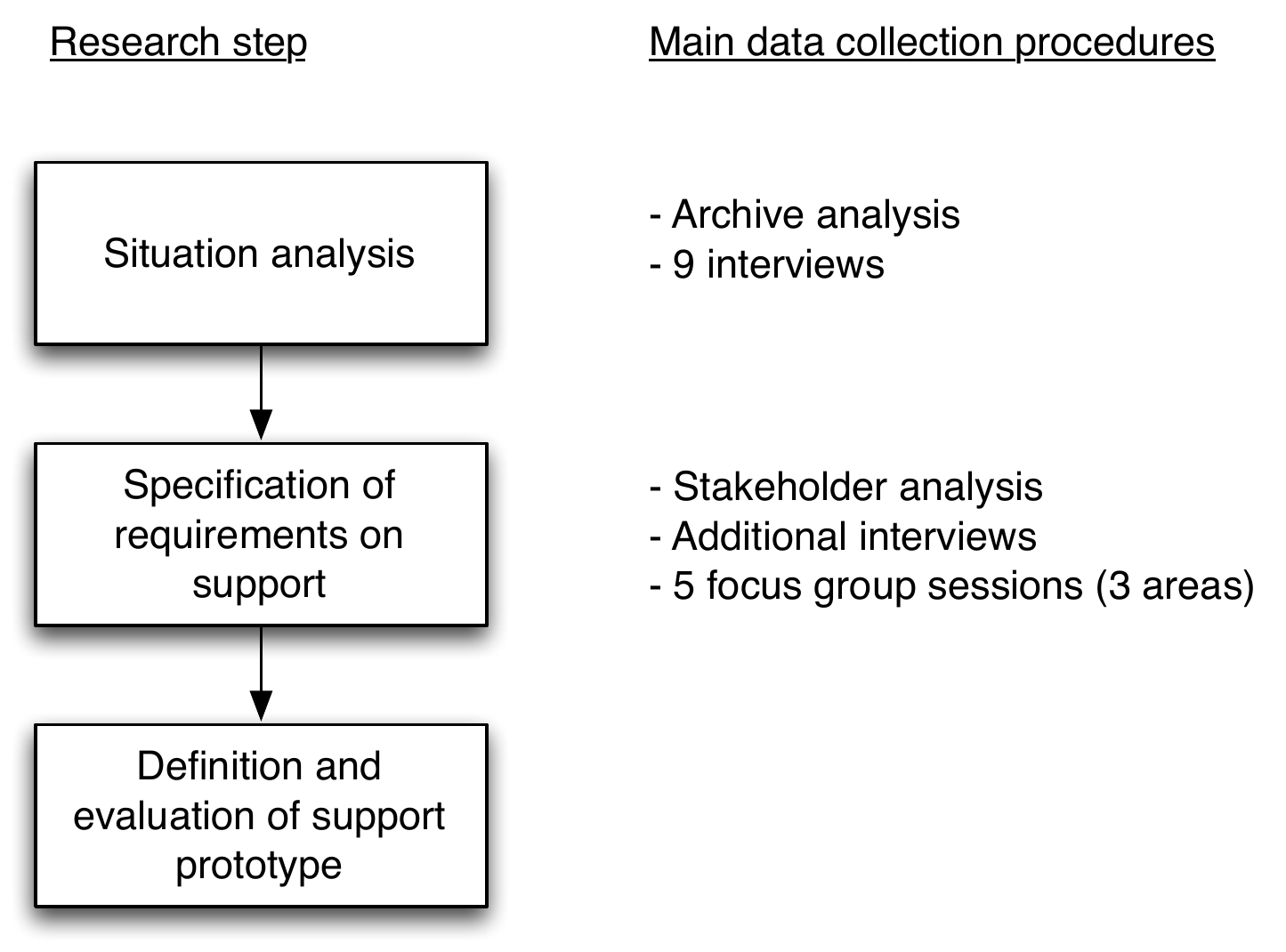}
\caption{Overview of data collection and analysis}
\label{fig:methodology}
\end{center}
\vspace{-20pt}
\end{figure}

The data collection and analysis was carried out as outlined in Figure~\ref{fig:methodology}, and described below. 

\subsection{Situation analysis}
A situation analysis was conducted with the objective to describe and somewhat explain the current situation at the company. 
The main goal was to gain an understanding of how work is conducted within the organization and can be seen as an observational part of the research.

Qualitative data was collected by studying documentation at the case company and by interviewing a sample group. 
The criteria for selecting people in this phase were that they should: there should be representatives from different areas, with different work tasks, to get a good variation of answers; have some work experience, within this or other companies; and be available for interviews. 


In total 9 individuals were interviewed which included 4 project managers/technical leaders/senior back-end developers, 2 senior front-end developer, 1 junior front-end developer, 1 junior back-end developer, and 1 service manager.

The interviews were semi-structured with about 20 questions prepared in advance. The interviews covered areas such as the experience, roles and responsibilities of the interviewee, the interviewees' view on how work is carried out today, the interviewees' opinion about the ideas of inner source at the
case company, and the interviewees' previous experience of new technology introduction. The opinion about the possibility of introducing inner source at the case company was of course an important part of the interviews.

All interviews were recorded with the permission of the participants. During the interviews, one of the authors acted as an observer, focusing on taking supporting notes. The notes were clarified directly after the interviews and written down in an interview summary.

The interview data was then analysed using an editorial approach (e.g.\ \cite{Runeson12}), meaning that the categories and statements for characterizing the reasoning in the interviews were not to a large extent predefined.

The framework presented by Stol \cite{Stol2011}, with some modifications to suit the company, were then used to evaluate the compatibility of the company to adopt inner source. 

\subsection{Specification of requirements on technical and practical support}
To define the technical infrastructure, related context and practices, a requirement specification was chosen because it is a natural approach within the software industry to describe a desired solution. The requirements specification is not presented here, although an overview of the domain can be found in Section \ref{OverviewTechnicalSolution}, and more information is provided in \cite{thesis}.

Several methods were used in order to elicit requirements from all levels of interest, e.g. stakeholder analysis, additional interviews, and a series of focus groups.

\subsubsection{Stakeholder analysis.}
A stakeholder analysis (e.g. \cite{Lauesen02}) was used to map all of the stakeholders and elicit their different areas of interest. It is important that everyone with a stake in the product gets to contribute their view, goals and wishes concerning both functional and non-functional requirements in order for the final product to get a corporate wide approval. 

Stakeholders were identified amongst developers, project- and service managers, team managers and corporate representatives. The analysis was based on material from the interviews held in the situation analysis, complemented by the focus group meetings and a longer interview with the case companys' former CTO.

\subsubsection{Focus groups.}
As it became clear early on in the case study that stakeholders had different opinions and priorities, this technique was considered appropriate. The incentive was to create an understanding between stakeholders in addition to identify problems and gather ideas and opinions in a structured manner \cite{Lauesen02}. The other objective of the focus groups \cite{Kontio04} was to elicit requirements for a substantial part for the proposed solution. Three areas with different themes were therefore identified on which the focus groups were based upon: Reuse of code and knowledge; Tools and functionality; Time, sales strategy and incentives.

With these themes the authors regarded to have covered all relevant aspects of the product. Several subtopics were then identified around which the discussions were held.

The focus groups were carried out in 1.5 hour sessions. Focus groups 1 and 2 were both split into two sessions, while focus group 3 was carried out in one single session. The sessions were moderated by one of the two first authors, whilst the other documented by audio recording and taking supporting notes.



Each session had a brief list of subtopics where participants were allowed to briefly describe bad experiences and focus more on the ideal usage and functions. 
Post-it notes were used by the participants to record their opinions, where considered appropriate by the authors. These were then collected by the moderator and presented for a joint discussion. The discussions also included different aspects of risk,
cost and benefits of the proposals. Where different opinions were present, a collective prioritization of the ideas was conducted and motivation to the priorities encouraged by the moderator. The sessions concluded with a summary by the moderator. 


\section{Results}
In this section the results are presented. 

\subsection{Situation analysis} \label{SituationAnalysis}
As recognized before, reuse today within the division is seen as insufficient. It mainly happens by a ''mouth-to-mouth'' spread of what has been developed before and where it can be found. That is, there is not sufficient systematized knowledge available on what software is available for reuse. Certain
modules and functions, which are commonly used, risk being re-developed. For example, one interviewee stated 

\begin{quote}
''We could benefit a lot from having our own demo site or basic platform, including common modules, that projects can be based on.''
\end{quote}

and

\begin{quote}
''I use standard modules that are needed in projects that I sometimes know that someone else has done in another project or that I have done myself in another project and thereby I can use it. In other cases, we are not aware of it. Especially when a new developer enters [a project] who has not been around for so long and do not know what is available.''
\end{quote}

This also raises potential for a common framework that can be used as a standard template in many of the projects. An apparent need for a platform facilitating reuse exist, since redundant work is conducted. Some solutions are however considered too customer specific in order to be reusable in other projects. Two other important aspects are time and budget. These two factors are tightly knit together. The time set for documentation is seldom used for this specific purpose. Transfer of knowledge in general is a subject that needs to be incorporated in the day-to-day work process in every project. There is little or no time between projects for project feedback and knowledge transfer. Time estimations are tight in order to win customer deals and chargeable coverage is of high priority, leaving limited time for internal improvements. 

Concerning standardized tools, a common set of collaborative tools are in place, which also is positive from an inner source perspective, including an application lifecycle management tool (TeamForge\footnote{\url{http://www.collab.net/products/teamforge}}) which is under evaluation. There is also an open discussions ongoing in the case company, and willingness to change exist, even if time is a restraining factor for internal improvements. It is possible to identify classes and functions for reuse, but the extraction may be very time consuming. Concerning maintenance, there are little or no time between projects and development projects are generally transferred to maintenance projects after acceptance test.

There are some aspects that require extra effort if the case company choose to work with inner source. 
Modularization of code is needed, which may require training. Also, requirements are project-specific and are mainly set at the start of the project, but also constantly evolving in each sprint. 
However, the most important issue is probably about code ownership. The customer is the owner of the code, which may result in constraints on what can be reused. Another issue is related to communication. Developers sit closely together and are unlikely to benefit from ``open'' communication. Communication with customers is desired to be closer and steered away from e-mailing.

\begin{table}[htbp]
\begin{center}
\label{SummaryChart}
\caption{Summary chart of findings from interviews in relation to inner source practises. Elements based on framework by Stol \cite{Stol2011}}
\scalebox{0.8}{
\begin{tabular}{|p{3cm}|p{12cm}|}
\hline
\textbf{Element} &\textbf{Findings from interviews}\\
\hline
\multicolumn{2}{|l|}{\textbf{Software product}} \\
\hline
Runnable software &Classes and functions can be identified from previous projects, but the extraction may be very time consuming.\\
\hline
Needed by several project groups &There is potential for a common framework that can be used in several projects. An apparent need for a platform facilitating reuse exist, since redundant work is conducted.\\
\hline
Maturity state of the software &Constantly evolving techniques and modules for customer-specific solutions. \\
\hline
Utility vs simplicity &Some solutions may be too specific for the project in order to reuse. \\
\hline
Modularity &Modularization of code is needed, which may require training.\\
\hline
\multicolumn{2}{|l|}{\textbf{Development practices}} \\
\hline
Requirement elicitation &Requirements are project-specific and are mainly set at the start of the project, but also constantly evolving in each sprint. \\
\hline
Implementation and quality control &Agile, sprint-driven development, planned per sprint. The level of competence in and knowledge of the process used varies. Senior developers review junior developers informally. Because of insufficient unit testing, quality can sometimes be an issue. Testing is to some extent ''bazaar-like'' and peer-testing is performed as much as possible. \\
\hline
Release management &Frequent releases, after each sprint. Customers are provided with prototypes. \\
\hline
Maintenance &Little or no time between projects. Development projects are generally transferred to maintenance projects after acceptance test.\\
\hline
\multicolumn{2}{|l|}{\textbf{Tools \& Infrastructure}} \\
\hline
Standardized tools &Common set of collaborative tools are in place, though older projects remain using older tools. Freedom to select tools locally.\\
\hline
Infrastructure for open access &Projects are archived in a traditional folder structure. A project platform for distributed development has been initiated and is under evaluation. \\
\hline
\multicolumn{2}{|l|}{\textbf{Organisation \& Community}} \\
\hline
Work coordination &Developers are assigned tasks. In order to control that the correct tasks are prioritized, developers may switch between projects. Better overview desirable.\\
\hline
Communication &Developers sit closely together and are unlikely to benefit from ''open'' communication. Communication with customers is desired to be closer and steered away from e-mailing.\\
\hline
Leadership and decision making &Discussions are considered open and inputs appreciated. Evangelists and/or core team needed to take responsibility for a common framework. \\
\hline
Motivation and incentives &A lack of time is the biggest concern. Attractiveness of the tool also considered a critical success factor.\\
\hline
Open culture &Open discussions and willingness to change exist, though time a restraining factor for internal improvements.\\
\hline
Management support &Budget constraints a concern. Chargeable occupancy important. Plan for incorporating activities related to reuse in the sales strategy needed.\\
\hline
\multicolumn{2}{|l|}{\textbf{Additional factors}} \\
\hline
Project feedback and knowledge sharing &More feedback on a division level desirable to improve knowledge sharing across projects. \\
\hline
Project initiation &Dependent on a few individuals because of expertise needed to set up projects. \\
\hline
Code ownership &The customer is the owner of the code, which may result in constraints on what can be reused. \\
\hline
\end{tabular}
}
\end{center}
\end{table}

\subsection{Overview of technical infrastructure}
\label{OverviewTechnicalSolution}
\begin{figure}
\begin{center}
\includegraphics[width=\textwidth]{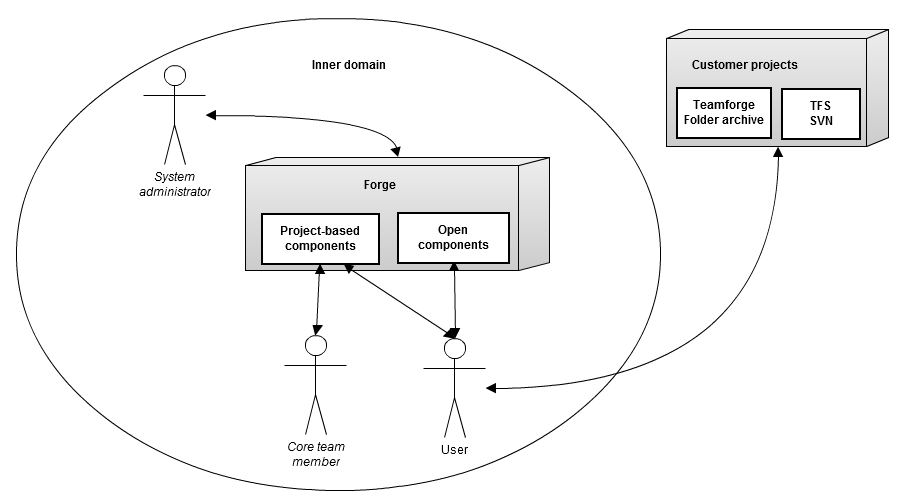}
\caption{Context Diagram}
\label{ContextDiagram}
\end{center}
\vspace{-20pt}
\end{figure}

This section presents the results from the specification of requirements and the definition of technical support for the case company. The domain for the technical solution consists of the forge, the users of the forge, and the system administrator within the studied division of the case company. The systems for the running customer projects, as well as the documentation of old projects, are outside the domain. The forge includes a component library with a component project view for each shared component, see Figure~\ref{ContextDiagram}.

\subsubsection{Components.}
All components are stored in a component library. 
A search function allows the user to find a component of interest. There are two types of components that can be shared, project-based components and open components.

Project-based components require administration by a core team that is responsible for the maintenance and development of the component. 
All users can access the components but the core team may restrict the user's rights to make any changes. 
Different types of components that have been identified as project-based components are Product, Framework and Templates. 
\begin{itemize}
\item \textbf{Product:} A basic solution that is ready to be sold to the customer, or to be customized. This type is a long term idea and though the forge provides the infrastructure to share this component, special requirements to support this type of component will not be further considered in this project.
\item \textbf{Framework:} A framework for the most commonly used modules. The framework would be used to have an initial set of modules that are reusable and can easily be implemented in a new project. A core team decides what goes into the framework and makes sure the framework is up to date and of the required quality. 
\item \textbf{Templates:} Documentation templates developed by a core team in order to facilitate the documentation process. 
\end{itemize}

Open components do not need any administration or anyone responsible for the development of the component. 
No quality or generalization requirements exist to share these components and the creator is not responsible for any maintenance or further development. 
Modules and Classes, and Knowledge Base are types of components identified as open components.
\begin{itemize}
\item \textbf{Modules and classes:} Modules and classes that have been used in projects. The size and complexity of these components may vary and may be more or less suitable to reuse.
\item \textbf{Knowledge bas:} User guides for commonly performed tasks, tutorials, lessons learned from previous projects, common errors etc.\ that could be helpful in future projects. 
\end{itemize}

\subsubsection{Users.}
The identified user types are all employees with access to the forge. They can be divided into three groups:
\begin{itemize}
\item \textbf{General user:} Developers reusing and contributing to components. The general users have full rights to open components and limited rights to project-based components.
\item \textbf{Core team member:} Project-based components have at least one core team member. The core team members have full access to its components and are responsible for maintenance, support and further development of that component.
\item \textbf{System administrator:} The system administrators is responsible for the technical support and maintenance of the forge. 
\end{itemize}

\subsubsection{Component project view.}
The component project view is unique and scalable, dependent of whether the component is open or project-based. Generally, users should be able to browse between the different versions of the component and read what changes that has been made by whom. Once the user has chosen a version of the component, a snapshot should be available for download. The user can then start investigating the component and see if it matches his expectations and requirements. If the user wishes to contribute, e.g report a bug or suggest a new feature this should be possible through an issue tracker or a by submitting a comment. The user can also choose to make the change by him or herself by accessing the code through the configuration management interface.

For project-based components specifically, it is up to the core team members to decide and configure how they want the component to be managed and accessed to other users. The creator of the component is automatically assigned the role as a core team member. This individual can add additional users to the core team when appropriate. The component can be free for everyone to see and download or may be restricted to an individual or groups of users. This is to allow different levels of core team responsibility depending on the criticality of the component.

Depending on the core team's preferences and the conditions of the component, different processes can be used. A general option in open source development and closely linked to project-based inner source is that users have free access to the communal component project view where they can share information, communicate and access the repository but with reading rights exclusively. They are free to make change/feature requests and bug reports, but all changes and features made by themselves are sent in as deltas via the configuration management interface. These deltas can then be reviewed by the core team and either be sent back or implemented into the component repository. With this way of working, core team members have full control of what goes into the component and that re-factorizations needed are done properly.

\subsection{Work process}
The level to which inner source is adopted depends on the organization and can be done in various ways. This Section aims to provide an organizational context to the technical solution and discuss how to adopt inner source practices within the studied division, based on the situation analysis, the requirement elicitation and findings from the literature. 

\subsubsection{Development practices.}
As described in Section \ref{OverviewTechnicalSolution}, the domain does not include the customer projects. Hence, the introduction of inner source, as proposed here, will have a low impact on the development practices used. Collaborative development is used to some extent and under improvement with the introduction of TeamForge (TF), which would benefit the implementation process of the forge.

It was revealed in the situation analysis, that the quality of the code is an issue from time to time. Hence, it is relevant to take advantage of the quality benefits associated with inner source. Clear visualization of ratings and issues as well as test- and review results for each components is therefore an important aspect. That way, quality can be improved both directly on the specific component and indirectly by allowing developers to gain skills through the identification of errors. 
A drawback of ratings, anticipated by the team, is that it can be misleading. Users who have tested or reviewed the component may have done this to different extents why their conception of the component may vary. This is why additional information such as tagging, rating, comments and descriptions are considered important so that whole experiences are reflected and potential users can get a fair comprehension of the component. 

Alignment between the different projects is important to the team. A framework is considered to improve alignment of the development from project to project through a communal set of components, optimizing the initiation process and facilitating for developers to enter a new project.

\subsubsection{Organization and community.}
To complement the technical aspect of adopting inner source in the division, organizational aspects as well as community building are discussed in this section. Though the general organizational structure does not need to change, some effort on all levels is required to adapt to the new conditions and create business value from the initiative.

With the solution proposed, a volunteer approach to the work coordination related to open and project-based components will be used, in contrast to assigned tasks. To encourage contributions, sharing open components does not require any further responsibility from the creator, nor a leadership or decision making structure. The motivation is to limit the responsibility of the creator, decrease dependency on individual developers and enable assets to be highly dynamic. 

Regarding the project-based components, there is a need for coordination and management since these components are of a more business critical character. Depending on the complexity of the component, the amount of resources needed by the core team may vary. For a critical asset such as a framework, the core team members need to have deep technical knowledge as well as an understanding of the business- and delivery models. 

An incentive and motivational structure was identified in literature as essential for users to be attracted to the forge. From a management perspective, there is a wish to acknowledge the competent developer and the platform could be used as a tool for doing so. However, it is required to put some thought into what is being measured, to prevent that rewards, if any, are not misleading, nor cause negative implications for contributing. Additionally, it demands of the technical solution to provide these measures on an individual developer level in a simple manner for the manager to extract. 

The individual developer may be motivated to contribute by the rewards and acknowledgement of management, but motivation is also a highly cultural matter that needs to be incorporated in the working environment. Developers as well as managers and technical leaders, encouraging each other to contribute and to use the forge, foster awareness and integration of the solution in the day-to-day work. The need for an ``evangelist'' has been described in both literature \cite{Dinkelacker02} \cite{Sabbah05} and the situation analysis, Section \ref{SituationAnalysis} as essential for the success of an inner source initiative. This important role is hence to be chosen carefully and early on in the initiation process in order to push the projects forward.

As described by Wesselius, \cite{Wesselius08}, one of the limiting external factors of inner source development is overall profitability. That is, that the group should not optimize its own profits at the expense of the company's total profitability. By constantly retaining an awareness of what exists on the forge and the quality of it, individuals with responsibility for sales can adapt their estimates to presumptive customers. This cross-divisional dependency calls for a communication and discussion between the different internal stakeholders. Planning and development of the assets on the forge is of communal interest since it benefits the whole company.

\section{Expected implications}
It is found that the investigated case has good potential for adopting inner source. The division has plenty to gain by adopting the open source practises on which inner source is based, e.g.

\begin{itemize}
	\item Improved reuse of code and solutions to complex problems \cite{Lindman08}, \cite{Dinkelacker02}, \cite{Sabbah05}, \cite{Gurbani06}, \cite{Gurbani10}, \cite{Wesselius08}, \cite{Stol2011} \cite{Morgan11}
	\item Improved quality of code and general level of knowledge amongst developers \cite{Riehle09}, \cite{Sabbah05}, \cite{Martin07}, \cite{Gurbani06}, \cite{Gurbani10}
	\item Creation of a framework to standardize and shorten initiation process of new projects  \cite{Dinkelacker02}, \cite{Martin07}
	\item Better visibility and spread of information and knowledge \cite{Lindman08}, \cite{Dinkelacker02}, \cite{Sabbah05}, \cite{Stol2011}, \cite{Morgan11}
	\item Higher margins for tender processes \cite{Martin07}, \cite{Gurbani06} \cite{Gurbani10}, \cite{Wesselius08}, \cite{Stol2011}, \cite{Morgan11}
\end{itemize}

Emphasis should be on these benefits in the planning and evaluation processes to keep focus on what the organization want to achieve with the system. 

In order to receive these benefits and prosper from them, several challenges has to be overcome. Stol et al.\ \cite{Stol11} have identified numerous challenges, and there are some of special concern to the case company, e.g.
uncertainty about the quality of a component, 
awareness of the library content,
finding the right component,
motivating developers to contribute,
commitment in terms of devoting time for the developers to contribute, 
and
costs of creating modular and generic components.



\section{Validity}

The interviews were analysed based on the compatibility framework developed by Stol in \cite{Stol2011}. The framework has not been evaluated before by others than the author, but was developed based on an extensive literature study and case study within the subject performed by Stol et al.\ \cite{Stol10} \cite{Stol11}. For this case study it was considered a valuable tool for structuring the findings from the interviews from an inner source perspective. 

Since this is an individual case study, it can not be established if the technical solution and recommendations are applicable to other case companies. Additionally, the solution proposed has yet not been implemented, nor evaluated. The authors consider the solution to be an option for similar size of development teams, being a small company or part of a larger company that want to experiment with the adoption of inner source on a smaller scale before making significant investments. Evaluation of the solution would be needed in order to investigate what challenges the solution truly impose as well as the benefits similar organizations can expect to gain.

The main measures taken to improve the validity (e.g. \cite{Runeson12}) in the case study can be summarized as follows. 

\emph{Prolonged involvement} was achieved since the two main authors spend most of their working time at the premises of the case company during a time period of about 4 months. The fact that the researches spent so much time there and that they were able to build a network of interested engineers in the  organization made it possible to have relevant discussions about the results, and also to get access to required data.

\emph{Peer debriefing}, meaning that fellow researchers comment on the results was achieved by having the third author reviewing the findings and research methodology during the research without being actively involved in the day-to-day data collection. 

\emph{Member checking}, in this case meaning that engineers at the
case company can review findings were achieved by having regular discussions about the results with the members. In particular there were one person who acted as contact person and main discussant part during the study.

\emph{Audit trails} were achieved by recording all interviews and
taking extensive notes during data collection phases.

\section{Conclusion}
\label{conclusions}
Several potential benefits and expected challenges for the case company have been identified. One of the main advantages of inner source, as perceived by the case company, is the possibilities it offers for reuse of code and other artefacts. 

In order to address the challenges seen in introducing inner source, this case study has proposed a technical infrastructure, presented in form of a requirement specification, together with an adapted working process and organizational context. The infrastructure forms a collaborative platform where knowledge and code can be shared in form of components, and where people can interact according to the principles of inner source. 

Two types of components were identified in order to address the types of data which the case company wishes to share internally. Project-based components which are, to some extent, to be seen as business critical and demands supervision by a core team. This can be related to the concepts of project-based inner source as identified by Gurbani et al. \cite{Gurbani10}. The other type, open components, relates in some parts to the concept of infrastructural inner source. It can also be compared to a combination of a knowledge base and a code snippet library. This type of component can in general be anything of general interest and creator is not obligated to any support or maintenance of it. 

During this research the framework of Stol \cite{Stol2011} was used as a framework in the analysis. We can conclude that this framework was useful for us during this purpose, and we believe that it includes relevant factors. It can also be noticed that the concept of project-based and open components \cite{Gurbani10} is useful when formulating the support needed in the organization. 

The division studied within the case company possesses potential for the application of inner source and if successfully applied, it can bring several rewards to the organisation by optimizing its resources. An eventual future implementation is yet to be studied.

Since this study has focused on one case, it cannot be generalized to other organization by default. Many of the findings though, can be of value to other cases where smaller teams and organizations are investigating the opportunities to introduce inner source, which is an area for future research. 

\subsubsection*{Acknowledgments.} The authors would like to express their gratitude to all the people at the case company for supporting the research and participating in the study. This work was partly funded by the Industrial Excellence Center EASE - Embedded Applications Software Engineering, (\url{http://ease.cs.lth.se}).

\bibliographystyle{plain}
\bibliography{MUNThesisBibTeX}

\end{document}